\documentclass[pra,aps,twocolumn,nopacs,superscriptaddress,nofootinbib,longbibliography]{revtex4-1}

\usepackage{amsmath}  \usepackage{amssymb}  \usepackage{amsfonts}  \usepackage{bm}  \usepackage{bbm}   \usepackage{bbold}  \usepackage{braket}  \usepackage{color}  \usepackage{comment}  \usepackage{dcolumn}  \usepackage{enumerate}  \usepackage{epsfig}  \usepackage{gensymb}  \usepackage{graphicx}  \usepackage{indentfirst}  \usepackage{lmodern}  \usepackage{mathrsfs}  \usepackage{mathtools}  \usepackage{psfrag}  \usepackage{pst-all}  \usepackage{soul}  \usepackage{xcolor}
\usepackage{float}
\usepackage[colorlinks,linkcolor=blue,citecolor=blue,urlcolor=blue,hyperindex,driverfallback=dvipdfm]{hyperref}  \usepackage[T1]{fontenc}

\def\ii{{\rm i}}  \def\ee{{\rm e}}
  \def\kB{{k_{\rm B}}}
  
                              \def\kb{{\bf k}}        \def\pb{{\bf p}}              
          \def\eh{\hat{\bf e}}     
  \def\eps{\epsilon} 
\def\Hh{\hat{\mathcal{H}}}  \def\Hint{\Hh_{\rm int}}  
\def\ah{\hat{a}}  \def\bh{\hat{b}}  \def\ahd{\ah^{\dagger}} \def\bhd{\bh^{\dagger}}

\def\Om{\Omega}
\def\eps{\varepsilon}

\begin{document}

\title{Rotational Vacuum Friction of Nonabsorbing Particles
}

\author{F.~Javier~Garc\'{\i}a~de~Abajo}
\email{javier.garciadeabajo@nanophotonics.es}
\affiliation{ICFO-Institut de Ciencies Fotoniques, The Barcelona Institute of Science and Technology, 08860 Castelldefels (Barcelona), Spain}
\affiliation{ICREA-Instituci\'o Catalana de Recerca i Estudis Avan\c{c}ats, Passeig Llu\'{\i}s Companys 23, 08010 Barcelona, Spain}

\author{Alejandro~Manjavacas}
\affiliation{Instituto de Qu\'{\i}mica F\'{\i}sica Blas Cabrera (IQF), CSIC, 28006 Madrid, Spain}

\begin{abstract}
A nonabsorbing particle rotating in vacuum can lose angular momentum only by converting mechanical energy into electromagnetic radiation. Here, we develop a quantum theory of rotational vacuum friction for small lossless particles and show that axial symmetry qualitatively changes the leading dissipation channel. At zero temperature, the frictional torque scales as $M\propto\Om^7$ with rotation frequency $\Om$ in anisotropic particles due to the emission of correlated photon pairs whose frequencies sum to $2\Om$, while a contribution to the torque linear in $\Om$ is found at finite temperature. In contrast, axisymmetric particles are protected against photon-assisted friction regardless of temperature.
\end{abstract}
\date{\today}
\maketitle

\section{Introduction}

Quantum fluctuations of the electromagnetic vacuum exert noncontact forces and torques on neutral bodies in motion, a phenomenon generally known as vacuum friction \cite{SPS98,GK98,KG99,VP07,VP11,IBH16,JLD15,BMM01,WDT16,B12_2,PH13,BSB10}. An important manifestation is the dynamical Casimir effect, in which mechanical motion is converted into real photons~\cite{M1970_3,L94,LJR96,WJP11,LPH13,D10}. Another is friction arising between sliding planar surfaces separated by a gap, which has attracted sustained theoretical attention \cite{P97,P98_2,KEK11,MGK13,MJK14,WLL14}. Pendry \cite{P98_2} established a direct connection between dynamical Casimir emission and noncontact friction by showing that two transparent half-spaces sliding relative to one another dissipate energy through the creation of correlated photon pairs, one in each medium. This mechanism constitutes the lowest-order nonvanishing dissipation channel and becomes possible when the relative velocity exceeds the speed of light in the sliding media. The same multiphoton structure also appears in the rotating-particle system studied here. Other realizations include accelerated mirrors and oscillating cavities \cite{CF1984,L05_2}, as well as neutral atoms moving through inhomogeneous fields or orbiting one another \cite{B10_3,B10_4,BL15,LDJ12}.

A sphere spinning with angular velocity $\Om$ at zero temperature has been predicted to experience a vacuum-friction torque, radiating electromagnetic energy at the expense of mechanical rotation \cite{paper157,paper166,paper199,MJK12,MGK13_2,paper172,paper289,paper336,S14,LS16,RDH18,AXB18}, provided that the material is absorptive at optical frequencies below $\Omega$. For such an absorbing particle, the dominant channel is first order in the light--matter coupling, describing how one quantum of rotational energy $\hbar\Om$ is converted into a photon of frequency $\omega$ and an internal excitation of frequency $\eps$, subject to the energy constraint $\omega+\eps=\Om$ \cite{paper166}. The resulting zero-temperature torque scales as $\Om^5$ and can be written as a weighted integral of the absorptive part of the dipolar polarizability ${\rm Im}\{\alpha(\omega)\}$. Because the elementary process is linear in the light--matter coupling, the fluctuation--dissipation theorem (FDT) can be used to obtain this result \cite{paper157}, in agreement with a fully quantum treatment of the problem \cite{paper166}.

A natural question then arises, what becomes of vacuum friction when the particle is unable to absorb within the spectral range relevant to the rotational dynamics? More precisely, consider the regime in which $\Om$ lies below the optical band gap of the material in the particle. Then, the single-photon channel responsible for the leading-order torque in absorbing particles is no longer available, and rotational friction can only occur through higher-order processes that cannot be described through the FDT. A semiclassical treatment of rotating asymmetric particles indicates that photon emission is produced as a manifestation of the dynamical Casimir effect \cite{P05,P06,MBM26}, but a detailed quantum-mechanical analysis with a conclusive result for the friction torque in such a scenario is still lacking.

In this article, we show that nonabsorbing particles rotating in vacuum and lacking axial symmetry experience a frictional torque arising from multiphoton emission, thereby converting mechanical energy into radiation. The process leaves the particle's internal state unchanged while removing an even number of angular momentum quanta from the rotor. To the lowest order, the emission consists of a pair of photons with frequencies $\omega$ and $2\Om-\omega$, where $0<\omega<2\Om$. Physically, this mechanism constitutes the vacuum analogue of degenerate parametric down-conversion~\cite{MW95}, with the mechanical rotation itself providing the effective pump. Through a fully quantum-mechanical description of the rotor, the electromagnetic field, and the material excitations, we identify two-photon emission as the leading dissipation channel for such asymmetric particles, yielding a torque scaling as $\Om^7$ with rotation frequency $\Om$. In contrast, for nonabsorbing axisymmetric particles, including spheres, all photon-emission contributions vanish identically at any temperature, leaving us with no torque at zero temperature and a marginal torque produced by single-photon absorption above the band gap at finite temperature. This symmetry-controlled hierarchy of multiphoton processes defines the genuine zero-temperature quantum regime, whose finite-temperature crossover we also discuss.

\begin{figure}[htbp]
\centering\includegraphics[width=1.0\linewidth]{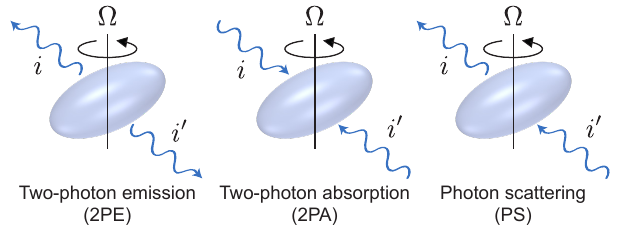}
\caption{A small nonabsorbing asymmetric particle rotates about the $z$ axis with angular velocity $\Om$. The particle can lose angular momentum through two-photon emission (2PE), two-photon absorption (2PA), and photon scattering (PS), all of which contribute to rotational friction. In particular, 2PE can occur at zero vacuum temperature, whereas 2PA and PS require a finite thermal photon population.}\label{Fig1}
\end{figure}

\section{Theoretical model}

We consider a particle with moment of inertia $I$ rotating about the $z$ axis with angular velocity $\Om$ in vacuum at temperature $T_0$ (Fig.~\ref{Fig1}). The particle supports bosonic internal excitations of frequencies $\eps_j$, assumed to lie well above the rotational and thermal scales ($\eps_j\gg 2\Om$ and $\eps_g\gg\theta_0\equiv2\pi\kB T_0/\hbar$, where $\eps_g={\rm min}\{\eps_j\}$ is the internal gap frequency). We use the basis $\ket{m,\{k_j\},\{n_i\}}$, where $\ket{m}$ is a rotational state with angular dependence $\ee^{\ii m\varphi}$, $\{k_j\}$ denotes internal-excitation occupations, and $\{n_i\}$ labels photon occupations in the surrounding electromagnetic vacuum \cite{paper166}. In the small-particle limit ($\Om,\eps_j\ll c/a$), the particle--radiation coupling is described within the electric-dipole approximation by the interaction Hamiltonian
\begin{align}\label{Hint}
\Hint=\sum_{ij}\sqrt{\frac{2\pi\hbar\omega_i}{V}}\;
\eh_i\!\cdot\!\pb_j\,(\ahd_i+\ah_i)(\bhd_j+\bh_j),
\end{align}
where $\eh_i$ are real unit polarization vectors, $V$ is the quantization volume, $a_{i}$ and $a_i^{\dagger}$ denote photon annihilation and creation operators, and $b_l$ and $b_l^{\dagger}$ are ladder operators associated with the internal excitation $j$. Photons couple to the lab-frame transition dipoles $\pb_j$, which are obtained from the real body-frame dipoles $\pb'_j$ by rigid rotation as $p_{j,x}=p'_{j,x}\cos\varphi-p'_{j,y}\sin\varphi$ and $p_{j,y}=p'_{j,x}\sin\varphi+p'_{j,y}\cos\varphi$. We do not consider particle-shape-dependent noninertial-force effects in this transformation \cite{paper336}, as they do not affect the main results discussed here. We omit polarization along $z$, which does not produce friction under the assumption that it can be uncoupled from polarization in the $x-y$ plane. From Eq.~(\ref{Hint}), for zero occupation numbers, the matrix elements that change the rotational quantum number are $\bra{m,0,0}\Hint\ket{m\pm1,1_j,1_i}=\Delta^{\pm}_{ij}$ with
\begin{align}\label{Dij}
\Delta^{\pm}_{ij}=\sqrt{\frac{\pi\hbar\omega_i}{2V}}\, (p'_{j,x}\mp\ii p'_{j,y}) (e_{i,x}\pm\ii e_{i,y}),
\end{align}
corresponding to a change $m\to m\pm1$ in the rotational quantum number, accompanied by the creation or annihilation of a photon $i$ and a particle excitation $j$.

\section{Rotational friction of nonabsorbing particles}

Starting from a rotational state $m_0$, a first-order transition would have to satisfy the energy-conservation condition $\pm\omega_i\pm\eps_j=\Om$, with $\Om=\hbar m_0/I$. Such processes are therefore forbidden by the large internal gap $\eps_g$. The leading allowed contribution for an anisotropic nonabsorbing particle is second order: the particle emits or absorbs two photons, it scatters one photon while returning to its internal ground state and changing its angular momentum by $-2\hbar$, $2\hbar$, and $\pm2\hbar$, respectively (Fig.~\ref{Fig1}). Two-photon emission (2PE) can occur at zero temperature, while two-photon absorption (2PA) and photon scattering (PS) require a finite photon population. Applying Fermi's golden rule to these second-order processes and averaging over the thermal photon occupations, we calculate transition rates $\Gamma_{2PE}$, $\Gamma_{2PA}$, and $\Gamma^\pm_{PS}$, where the signs $\pm$ in the PS terms label scattering processes that increase or decrease angular momentum. The corresponding torque is $M=2\hbar(-\Gamma_{\rm 2PE}+\Gamma_{\rm 2PA}+\Gamma^+_{\rm PS}-\Gamma^-_{\rm PS})$. Converting photon-mode sums according to $\sum_i\to V(2\pi)^{-3}\sum_\sigma\int d^3\kb$, and using $\sum_\sigma\int d^2\Omega_\kb\,|e_x\pm\ii e_y|^2=16\pi/3$ to sum over photon polarizations and directions of propagation, we obtain (see Appendix~\ref{appendixA})
\begin{widetext}
\begin{align}\label{M1}
M=-\frac{4\hbar}{9\pi c^6}\, \bigg\{&
\int_0^{2\Om} d\omega\,\omega^3(2\Om-\omega)^3 \big[n_{T_0}(\omega)+n_{T_0}(2\Om-\omega)+1\big]\, g(\omega-\Om)
\\\nonumber
+&\int_0^\infty d\omega\,\omega^3(2\Om+\omega)^3 \big[n_{T_0}(\omega)-n_{T_0}(2\Om+\omega)\big]\, g(\omega+\Om)
\bigg\},
\end{align}
\end{widetext}
with
\begin{align}\label{g1}
g(\omega)=\big|\alpha_{xx}(\omega)-\alpha_{yy}(\omega)\big|^2+4\big|\alpha_{xy}(\omega)\big|^2.
\end{align}
Here, $n_{T_0}(\omega)=[\exp(\hbar\omega/\kB T_0)-1]^{-1}$ is the Bose--Einstein distribution at the vacuum temperature $T_0$. Importantly, Eq.~(\ref{M1}) is written entirely in terms of the particle polarizability tensor, which we have identified from its Cartesian components $\alpha_{ab}(\omega)=\frac1{\hbar}\sum_j p'_{j,a}p'_{j,b}[(\eps_j+\omega)^{-1}+(\eps_j-\omega)^{-1}]$ with $a,b\in{x,y}$ (i.e., the standard form obtained from the Hamiltonian in Eq.~(\ref{Hint}) \cite{CT1984,B12_2}). The polarizability is real over the spectral range considered here because the particle is nonabsorbing. In addition, for $\Om\ll\eps_g$, we may approximate $g(\omega\pm\Om)\approx g(0)\equiv g_0$, allowing $g_0$ to be taken outside the frequency integrals in Eq.~(\ref{M1}).

\begin{figure}[htbp]
\centering\includegraphics[width=1.0\linewidth]{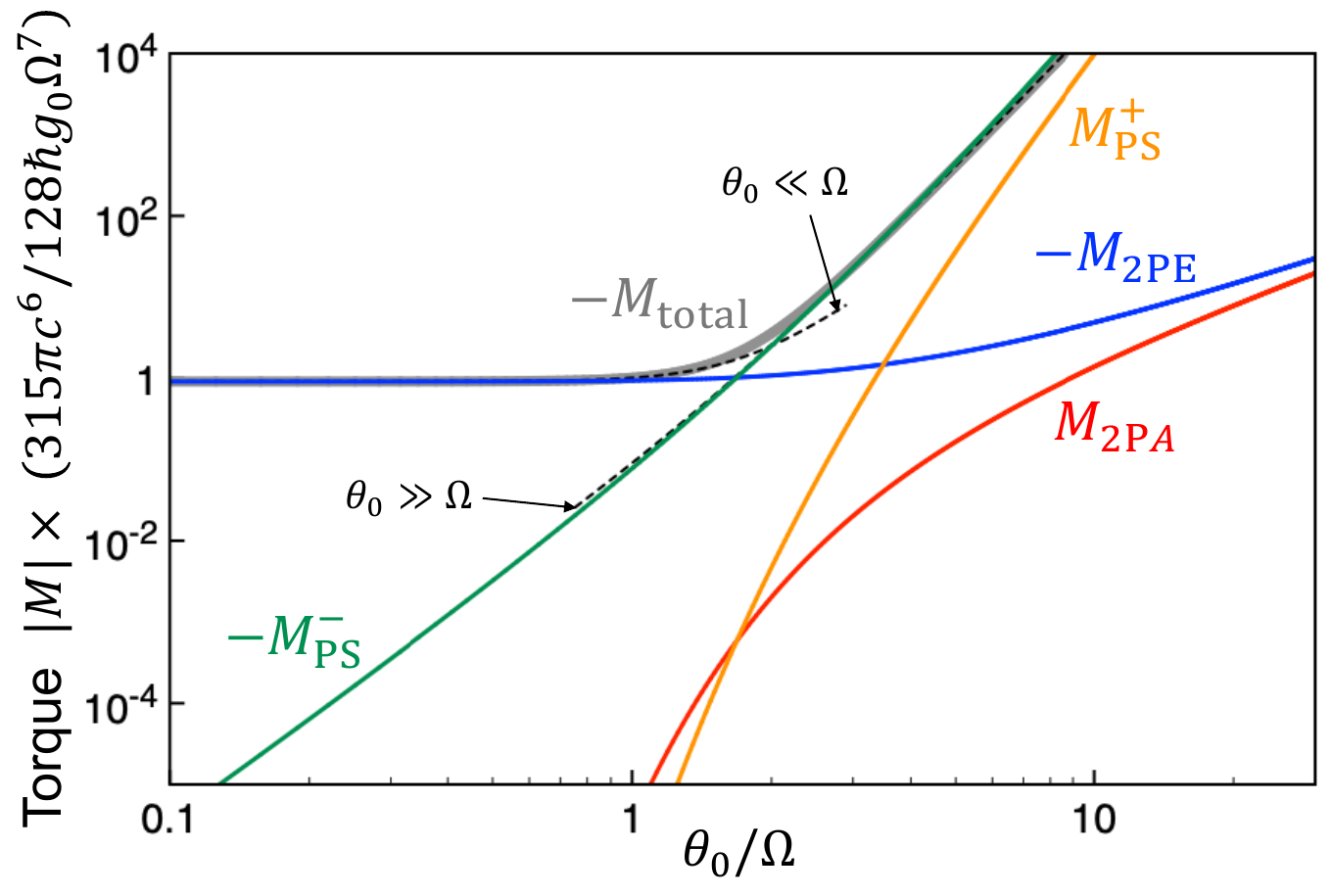}
\caption{Temperature and velocity dependence of rotational friction in nonabsorbing particles. We show different contributions to the torque $M_s$ arising from the processes depicted in Fig.~\ref{Fig1} and dependent only on the $\theta_0/\Omega$ ratio, where $\theta_0=2\pi\kB T_0/\hbar$ is the thermal frequency. The $s=\,$PS contribution is separated into terms associated with $\pm\hbar$ momentum transfers. The total torque (gray curve) is compared with the high- and low-temperature limits (broken curves).}\label{Fig2}
\end{figure}

In the low-temperature regime ($\xi\equiv\Om/\theta_0\gg1$), Eq.~(\ref{M1}) gives
\begin{subequations}\label{asym}
\begin{align}\label{largexi}
M=-\frac{128}{315\pi}\frac{\hbar\Om^7g_0}{c^6} \left[1+\frac{7}{64}\,\xi^{-4}+\mathcal O(\xi^{-5})\right]
\end{align}
(see Appendix~\ref{appendixB}). Thus, at zero temperature, the rotational vacuum friction scales as $\Om^7$. In the opposite limit ($\xi\ll1$), we find
\begin{align}\label{smallxi}
M=-\frac{2}{189\pi}\frac{\hbar\theta_0^6\Om g_0}{c^6}\left[1+\frac{14}{5}\,\xi^2+\mathcal O(\xi^5)\right],
\end{align}
\end{subequations}
so that the leading high-temperature torque is linear in $\Om$ and scales with the sixth power of the temperature.

The contributions to the torque associated with the diagrams in Fig.~\ref{Fig1} are compared in Fig.~\ref{Fig2}, where photon scattering is separated into processes that accelerate ($M^+_{\rm PS}$) or decelerate ($M^-_{\rm PS}$) the rotation. At low temperatures, the torque is dominated by 2PE, whereas at high temperatures, losses associated with photon scattering produce the leading contribution. The asymptotic expressions in Eqs.~(\ref{asym}) agree well with the full calculation below and above $\theta_0\sim\Om$, respectively, as shown by the gray and broken curves in Fig.~\ref{Fig2}.

\begin{figure}[htbp]
\centering\includegraphics[width=\columnwidth]{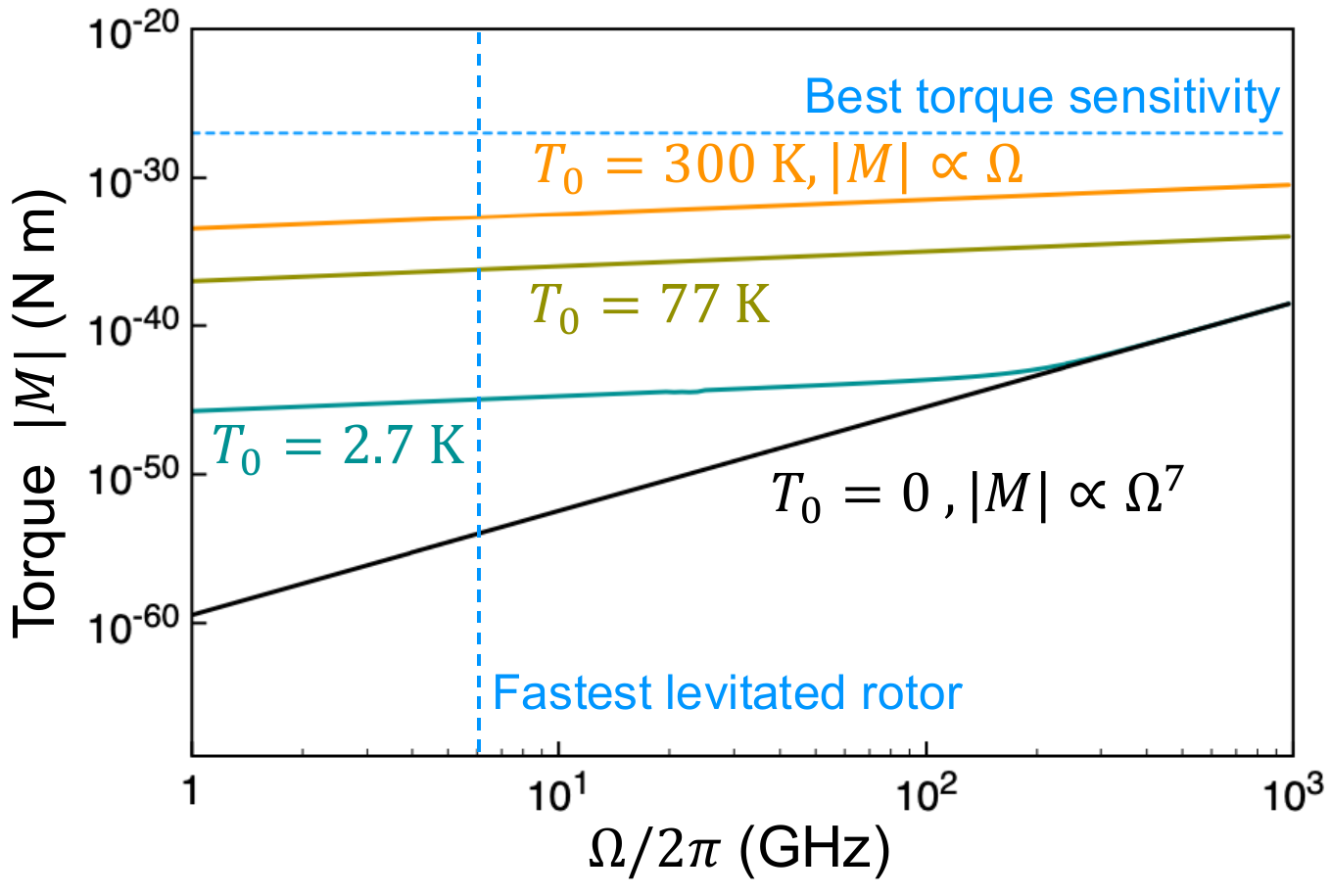}
\caption{Absolute strength of rotational vacuum friction in a nonabsorbing asymmetric particle. We plot the frictional torque as a function of rotation frequency $\Omega$ for a prolate diamond ellipsoid with semi-axes of 100~nm and 76~nm at different vacuum temperatures $T_0$ (see labels). The vertical and horizontal dashed lines indicate the fastest mechanical rotation reported to date ($\Omega/2\pi=6$~GHz \cite{JYR21}) and the measured torque sensitivity in levitated nanorotors ($\sim1\times10^{-27}$~N\,m \cite{AXB20}), respectively.}\label{Fig3}
\end{figure}

To place these results in the context of state-of-the-art levitated optomechanics, we consider a prolate diamond ellipsoid, which is transparent from microwave to optical frequencies \cite{P1985} and has a static permittivity of 5.7 \cite{DNC98}). We take semi-axes of 100~nm and 76~nm, yielding $g_0=4.8\times10^{-7}~\mu$m$^6$ for rotation about a short axis, and consider the highest reported rotation frequency, $\Omega/2\pi=6$~GHz \cite{JYR21}. Plugging the analytical expression of the static polarizability \cite{O1945} into  Eq.~(\ref{M1}), we find $|M|\sim2\times10^{-33}$~N\,m at $T_0=300$~K [from Eq.~(\ref{smallxi}), since $\theta_0/\Om\gg1$], about six orders of magnitude below the record sensitivity of levitated torque sensors \cite{AXB20}, and corresponding to spin-down times $I\Omega/|M|\sim10^{12}$~s (see Appendix~\ref{appendixC}). Transparent nanorotors are therefore essentially immune to vacuum friction, in sharp contrast to absorbing particles \cite{paper157}. The two-photon emission channel is thus irrelevant as a source of decoherence in proposed tests of quantum rotation with levitated particles \cite{SHK21}, rendering quantum rotation states robust for storing and processing quantum information.

\section{Frictionless rotation of nonabsorbing symmetric particles}

The tensor combination in Eq.~(\ref{g1}) shows that the two-photon torque vanishes for particles that are symmetric about the rotation axis, for which $\alpha_{xx}=\alpha_{yy}$ and $\alpha_{xy}=0$. This cancellation is the rotational analogue of the angular-momentum selection rule that forbids two-photon decay between atomic states of zero angular momentum when the dipole transitions to the intermediate manifold are isotropic. In both scenarios, anisotropy is required to break the degeneracy and yield a nonzero amplitude.

Beyond second-order processes, the vanishing of the torque for axisymmetric particles persists when higher-order interactions are included, provided the system remains in the large-gap regime. Odd-order processes vanish because, within the dipolar interaction assumed in Eq.~(\ref{Hint}), they cannot return the particle to its internal ground state. In addition, even-order processes are composed of pairs of matrix elements such as $\Delta^\pm_{i,j}\Delta^\pm_{i',j}$, which describe the emission or absorption of photons $i$ and $i'$ accompanied by excitation and subsequent de-excitation of an internal mode $j$. These matrix elements contain factors $p'_{j,x}\pm\ii p'_{j,y}$ [see Eq.~(\ref{Dij})], where the sign is fixed by whether the elementary transition decreases or increases the rotational quantum number. A nonzero frictional torque requires both matrix elements to carry the same rotational sense, producing factors of the form  $(p'_{j,x}\pm\ii p'_{j,y})^2=p'^2_{j,x}-p'^2_{j,y}\pm2\ii p'_{j,x}p'_{j,y}$. However, for an axisymmetric particle, the transverse dipolar response is invariant under rotations in the $x-y$ plane. Consequently, after summing over $j$, the $x$ and $y$ contributions are equal in magnitude but opposite in sign, while the mixed term vanishes. Thus, every even-order process that returns the particle to its internal ground state cancels upon summation over the internal modes, and the total frictional torque remains zero. In addition, final states in which the particle is internally excited are likewise suppressed in the large-gap regime since reaching such states would require the internal excitation energy to be built up from several rotational-energy quanta of order $\Om$, whose relevant amplitudes contain products such as $\Delta^+_{i,j}\Delta^+_{i',j}$, which cancel after summing over $j$ by the same symmetry arguments used above.

Axisymmetric particles with a large band gap, well above the rotational frequency scale, can still experience rotational friction at finite temperature. This occurs because thermal photons have a small but finite occupation at frequencies above the gap $\eps_g$, enabling direct absorption by the particle and producing rotational friction. Using the analytical expressions for the torque $M$ and absorbed power $P^{\rm abs}$ given in Ref.~\cite{paper157} to first order in the particle--radiation interaction, the equilibrium condition $P^{\rm abs}=0$ yields a particle--vacuum temperature difference $T_1/T_0\approx1-5\pi\Om^2/\eps_g\theta_0$ to lowest order in $\theta_0/\eps_g$ and $\Omega/\eps_g$ assuming $\Omega\ll\theta_0$, while the corresponding torque reduces to
\begin{align}\nonumber
M\approx-\frac{8\hbar\eps_g^3\Om}{3\pi c^3}\,h(\eps_g^+)\,\ee^{-\hbar\eps_g/\kB T_0},
\end{align}
where $h(\omega)=(1/2)\,{\rm Im}\{\alpha_{xx}(\omega)+\alpha_{yy}(\omega)\}$ (see Appendices~\ref{appendixD} and \ref{appendixE}). This expression shows a linear dependence on $\Om$, together with the expected exponential suppression by the factor $\ee^{-\hbar\eps_g/\kB T_0}$.

\section{Concluding remarks}

We have shown that a small axisymmetric particle rotating in vacuum at zero temperature with angular frequency $\Om$ below its optical band gap $\eps_g$ experiences no frictional torque. This result, which identifies a dynamically stable equilibrium configuration, is protected by axial symmetry and persists to all orders in the particle--radiation interaction. At finite temperature, thermal photons above the gap produce a nonzero frictional torque that is linear in $\Om$, but exponentially suppressed with the gap-to-thermal-frequency ratio $\eps_g/\theta_0$. The situation changes qualitatively for particles lacking axial symmetry. Even at zero temperature, such rotors experience a finite frictional torque that scales as $\Om^7$. This torque originates from the spontaneous emission of frequency-entangled photon pairs satisfying $\omega_1+\omega_2=2\Omega$ and carrying a total angular momentum $2\hbar$ along the rotation axis. Therefore, the rotating particle acts as a mechanically driven source of entangled photon pairs, for which coincidence detection would provide a background-free experimental signature. Parametric implementations of rotating anisotropy in superconducting circuits, where dynamical Casimir emission has already been observed~\cite{WJP11,LPH13}, offer a promising route to emulate and enhance the predicted $\Om^7$ pair-emission process. Our results further suggest a hierarchy governed by discrete rotational symmetry: a particle with $N$-fold symmetry about the rotation axis can exchange angular momentum with the photon field only through multipoles of order at least $N$, leading to additional suppression of the frictional torque by powers of $\Omega R/c$, where $R$ is the particle size. Quantifying this hierarchy requires extending the present treatment to include higher-order multipolar responses of rotating particles.

\acknowledgments
This work has been supported in part by the European Research Council (101141220-QUEFES), the Spanish MICIU (PID2024-157421NB-I00 and Severo Ochoa CEX2024-001490-S), and the CERCA Program. A.M. acknowledges support from Grant PID2022-137569NB-C42 funded by MCIN/AEI/10.13039/501100011033 and by ERDF/EU, and Grant EIC24-1-17304 from the Programa Fundamentos FBBVA.

\appendix

\begin{widetext}

\renewcommand{\theequation}{A\arabic{equation}}
\renewcommand{\thesection}{A}
\section{Theory of rotational friction of a nonabsorbing particle}
\label{appendixA}

We consider a particle with moment of inertia $I$ rotating about the $z$ axis with angular velocity $\Om$ in vacuum at temperature $T_0$. The particle supports internal bosonic excitations of frequencies $\eps_j$ above a gap $\eps_g={\rm min}\{\eps_j\}$, which we assume to be larger than the rotation frequency ($\eps_g>\Om$). In addition, the frequency difference $\eps_g-\Om$ is taken to be large compared with the thermal frequency $\theta_0=2\pi\kB T_0/\hbar$ (i.e., $\eps_g-\Om\gg\theta_0$). We describe the rotational degree of freedom using a complete basis of states $\ket{m}$, labeled by the azimuthal quantum number $m$ and with angular dependence $\ee^{\ii m\varphi}$. The surrounding electromagnetic vacuum is described by photon-number states $\ket{\{n_i\}}$, where $n_i$ is the occupation of mode $i$ with frequency $\omega_i$. Likewise, $\ket{\{k_j\}}$ denotes the occupation-number states of the internal bosonic excitations. The noninteracting Hamiltonian of the uncoupled internal, rotational, and radiation degrees of freedom reads
\begin{align}\nonumber
\Hh_0=\sum_j\hbar\eps_j\bh^\dagger_j\bh_j+\sum_m\frac{\hbar^2m^2}{2I}\ket{m}\bra{m}+\sum_i \hbar\omega_i \ah^\dagger_i\ah_i,
\end{align}
where $a_i$ and $a_i^{\dagger}$ are photon annihilation and creation operators, while $b_j$ and $b_j^{\dagger}$ are ladder operators of the internal excitations.

Particle--radiation interaction is described through the Hamiltonian
\begin{align}\label{Hint}
\Hint=\sum_{ij} \sqrt{\frac{2\pi\hbar\omega_i}{V}}\;
\eh_i\!\cdot\!\pb_j\, (a_i^{\dagger}\!+\!a_i)\, (b_j^{\dagger}\!+\!b_j),
\end{align}
where $\eh_i$ are real unit polarization vectors and $V$ is the vacuum quantization volume. Crucially, photons in Eq.~(\ref{Hint}) couple directly to the lab-frame transition dipoles $\pb_j$, which are related through
\begin{align}\label{ppp}
\begin{split}
&p_{j,x}=p'_{j,x}\cos\varphi-p'_{j,y}\sin\varphi, \\
&p_{j,y}=p'_{j,x}\sin\varphi+p'_{j,y}\cos\varphi, \\
&p_{j,z}=p'_{j,z}
\end{split}
\end{align}
to the real body-frame dipoles $\pb'$, under the assumption of rigid rotation. We do not consider particle-shape-dependent the effect of noninertial (Coriolis and centrifugal) forces in this transformation because they do not affect the main results discussed here.

In terms of the internal excitation energies and transition dipoles, the $ab$ Cartesian component of the particle polarizability in the body frame takes the standard form \cite{CT1984,B12_2}
\begin{align}\label{alpha}
\alpha_{ab}(\omega)=\frac{1}{\hbar}\sum_jp'_{j,a}p'_{j,b}
\Big[\frac{1}{\eps_j+\omega+\ii0^+}+\frac{1}{\eps_j-\omega-\ii0^+}\Big]
\end{align}
within the retarded dipole-response formalism for a quantum system initially prepared in its ground state. This expression is used below to write the rotational torque in terms of the macroscopic polarizability.

The states $|m,\{k_j\},\{n_i\}\rangle$ constitute a complete basis to describe the particle--radiation system. From Eqs.~(\ref{Hint}) and (\ref{ppp}), the only nonzero transition matrix elements are those in which a single photon occupation number changes by $\pm1$, a single internal excitation number changes by $\pm1$, and the azimuthal quantum number changes by $\Delta m=\pm1$ or $0$. These matrix elements can be written in terms of the single-photon, single-excitation transitions
\begin{subequations}
\begin{align}\label{Dpm}
\Delta^\pm_{ij}
&=\langle m,\{0\},\{0\}|\Hint|m\!\pm\!1,\{1_j\},\{1_i\}\rangle=\langle m,\{0\},\{1_i\}|\Hint|m\!\pm\!1,\{1_j\},\{0\}\rangle
\\\nonumber
&=\langle m,\{1_j\},\{0\}|\Hint|m\!\pm\!1,0,\{1_i\}\rangle=\langle m,\{1_j\},\{1_i\}|\Hint|m\!\pm\!1,0,\{0\}\rangle
=\sqrt{\frac{\pi\hbar\omega_i}{2V}}\, \bigl(p'_{j,x}\mp\ii p'_{j,y}\bigr)\bigl(e_{i,x}\pm\ii e_{i,y}\bigr),
\\\label{D0}
\Delta^0_{ij}
&=\langle m,\{0\},\{0\}|\Hint|m,\{1_j\},\{1_i\}\rangle=\langle m,\{0\},\{1_i\}|\Hint|m,\{1_j\},\{0\}\rangle
\\\nonumber
&=\langle m,\{1_j\},\{0\}|\Hint|m,0,\{1_i\}\rangle=\langle m,\{1_j\},\{1_i\}|\Hint|m,0,\{0\}\rangle
=\sqrt{\frac{2\pi\hbar\omega_i}{V}}\, p'_{j,z}e_{i,z}.
\end{align}
\end{subequations}
In what follows, we ignore $\Delta^0_{ij}$ because it conserves $m$ and, therefore, does not change the rotational state of the particle. We further consider relatively symmetric particle configurations in which the transition dipole $\pb_j$ associated with each internal excitation is either parallel or perpendicular to $z$.

Starting from an initial rotational state $m_0\gg1$, we use Fermi's golden rule to compute inelastic transition rates. Each first-order scattering event changes the rotational energy by $\hbar^2[(m_0\pm1)^2-m_0^2]/2I\approx\pm\hbar\Om$ with $\Om=\hbar m_0/I$ and, therefore, imposes the energy-conservation condition $\pm\omega_i\pm\eps_j=\Omega$. We take $\Om>0$ without loss of generality. Under the assumption of a large internal absorption gap, first-order transitions cannot conserve energy, apart from those associated with the small thermal population of particle excitations and vacuum photons, which we neglect because $\eps_g-\Om\gg\theta_0$ (see above). The next potential contribution to friction is thus associated with second-order processes that return the particle to its initial ground state. In these processes, the rotational quantum number changes by $\Delta m=\pm2$, corresponding to an angular-momentum change $\pm2\hbar$. The torque $M$ is then obtained by multiplying the corresponding transition rates by this angular momentum transfer. Combining these ingredients, we find
\begin{subequations}\label{MMMM}
\begin{align}
M=M_{2PE}+M_{2PA}+M^+_{PS}+M^-_{PS},
\end{align}
where
\begin{align}
&M_{2PE}=-\frac{4\pi}{\hbar^3}\sum_{ii'}\big[n_{T_0}(\omega_i)+1\big]\big[n_{T_0}(\omega_{i'})+1\big]\,
\bigg|\sum_j\Delta^-_{ij}\Delta^-_{i'j}
\Big(\frac1{\eps_j+\omega_i-\Om}+\frac1{\eps_j+\omega_{i'}-\Om}\Big)\bigg|^2
\delta(\omega_i+\omega_{i'}-2\Om),
\\
&M_{2PA}=\frac{4\pi}{\hbar^3}\sum_{ii'}n_{T_0}(\omega_i)\,n_{T_0}(\omega_{i'})\,
\bigg|\sum_j\Delta^+_{ij}\Delta^+_{i'j}
\Big(\frac1{\eps_j-\omega_i+\Om}+\frac1{\eps_j-\omega_{i'}+\Om}\Big)\bigg|^2
\delta(\omega_i+\omega_{i'}-2\Om),
\\
&M^\pm_{PS}=\pm\frac{4\pi}{\hbar^3}\sum_{ii'}\big[n_{T_0}(\omega_i)+1\big]\,n_{T_0}(\omega_{i'})\,
\bigg|\sum_j\Delta^\pm_{ij}\Delta^\pm_{i'j}
\Big(\frac1{\eps_j+\omega_i\pm\Om}+\frac1{\eps_j-\omega_{i'}\pm\Om}\Big)\bigg|^2
\delta(\omega_i-\omega_{i'}\pm2\Om)
\end{align}
\end{subequations}
are the separate contributions associated with two-photon emission (2PE), two-photon absorption (2PA), and single-photon scattering (PS), respectively. In deriving these expressions, we have averaged the torque over the thermal population of vacuum photon states $\{n_i^0\}$, which gives rise to factors involving the Bose--Einstein distribution
\begin{align} \nonumber 
n_{T_0}(\omega)=\frac{1}{\ee^{\hbar\omega/\kB T_0}-1}
\end{align}
for vacuum photons at temperature $T_0$.

We evaluate Eqs.~(\ref{MMMM}) by replacing the sums over photon modes $i$ and $i'$ with integrals over
wave vectors $\kb$ and sums over polarizations $\sigma$ according to the prescription
\begin{align}\label{sumi}
\sum_i\longrightarrow
\frac{V}{(2\pi)^3}\sum_{\sigma}\!\int d^3\kb.
\end{align}
Since the polarization vectors in Eq.~(\ref{Dpm}) are independent of $j$, they can be factored out of the corresponding sums in Eqs.~(\ref{MMMM}). The remaining angular integrations over vector directions, together with the polarization sums, are then performed independently for each photon, yielding
\begin{align}\label{sumsigma}
\sum_\sigma\int d^2\Omega_{\kb}\; \big|(e_x\pm\ii e_y)\big|^2=\frac{16\pi}{3}.
\end{align}
These relations allow Eqs.~(\ref{MMMM}) to be written as
\begin{align}\nonumber
\left[\begin{array}{l} M_{2PE} \\ \\ M_{2PA} \\ \\ M^\pm_{PS} \end{array}\right]
&=\frac{4}{9\pi\hbar c^6} \int_0^\infty\omega^3d\omega \int_0^\infty\omega'^3d\omega'
\\\nonumber
\times&\left[\begin{array}{l}
-\big[n_{T_0}(\omega)+1\big]\big[n_{T_0}(\omega')+1\big]\,
\bigg|\sum_j\bigl(p'_{j,x}+\ii p'_{j,y}\bigr)^2
\Big(\frac1{\eps_j+\omega-\Om}+\frac1{\eps_j+\omega'-\Om}\Big)\bigg|^2
\delta(\omega+\omega'-2\Om)
\\ \\
n_{T_0}(\omega)\,n_{T_0}(\omega')\,
\bigg|\sum_j\bigl(p'_{j,x}-\ii p'_{j,y}\bigr)^2
\Big(\frac1{\eps_j-\omega+\Om}+\frac1{\eps_j-\omega'+\Om}\Big)\bigg|^2
\delta(\omega+\omega'-2\Om)
\\ \\
\sum_\pm(\pm1)\big[n_{T_0}(\omega)+1\big]\,n_{T_0}(\omega')\,
\bigg|\sum_j\bigl(p'_{j,x}\mp\ii p'_{j,y}\bigr)^2
\Big(\frac1{\eps_j+\omega\pm\Om}+\frac1{\eps_j-\omega'\pm\Om}\Big)\bigg|^2
\delta(\omega-\omega'\pm2\Om)
\end{array}\right],
\end{align}
which can be simplified by using the $\delta$ functions to carry out the $\omega'$ integral, leading to
\begin{align}\label{M22}
&\left[\begin{array}{l} M_{2PE} \\ \\ M_{2PA} \\ \\ M^+_{PS} \\ \\ M^-_{PS} \end{array}\right]
=\frac{4\hbar}{9\pi c^6}\times\left[\begin{array}{l}
-{\displaystyle\int_0^{2\Om}} d\omega\; \omega^3(2\Om-\omega)^3\, \big[n_{T_0}(\omega)+1\big]\,\big[n_{T_0}(2\Om-\omega)+1\big]\; g(\omega-\Om)
\\ \\
\;\;{\displaystyle\int_0^{2\Om}} d\omega\; \omega^3(2\Om-\omega)^3\, n_{T_0}(\omega)\,n_{T_0}(2\Om-\omega)\; g(\omega-\Om)
\\ \\
\;\;{\displaystyle\int_0^\infty} d\omega\; \omega^3(\omega+2\Om)^3\, \big[n_{T_0}(\omega)+1\big]\,n_{T_0}(\omega+2\Om)\; g(\omega+\Om)
\\ \\
-{\displaystyle\int_{2\Om}^\infty} d\omega\; \omega^3(\omega-2\Om)^3\, \big[n_{T_0}(\omega)+1\big]\,n_{T_0}(\omega-2\Om)\; g(\omega-\Om)\end{array}\right],
\end{align}
with
\begin{align}\label{gw}
g(\omega)=\big|\alpha_{xx}(\omega)-\alpha_{yy}(\omega)\big|^2+4\big|\alpha_{xy}(\omega)\big|^2,
\end{align}
where we have identified the polarizability using Eq.~(\ref{alpha}) and employed the relation $n_{T_0}(-\omega)=-n_{T_0}(\omega)-1$. Combining different terms in Eq.~(\ref{M22}), we obtain
\begin{align}\label{M2bis}
M=-\frac{4\hbar}{9\pi c^6}\;
\bigg\{&\int_0^{2\Om} d\omega\; \omega^3(2\Om-\omega)^3\, \big[n_{T_0}(\omega)+n_{T_0}(2\Om-\omega)+1\big]\; g(\omega-\Om)
\\\nonumber
+&\int_0^\infty d\omega\; \omega^3(2\Om+\omega)^3\, \big[n_{T_0}(\omega)-n_{T_0}(2\Om+\omega)\big]\; g(\omega+\Om)
\bigg\}.
\end{align}
At zero temperature, we readily obtain the result $M=-(8\hbar/9\pi c^6)\,\int_0^{\Om} d\omega\; (\Om^2-\omega^2)^3\, g(\omega)$.

Because the internal excitation frequencies are much larger than the rotation and thermal frequencies $\Om$ and $\theta_0$, we may set $g(\omega\pm\Om)\approx g(0)\equiv g_0$ [see Eqs.~(\ref{alpha}) and (\ref{gw})]. This corresponds to a response governed by the static polarizability tensor, which enters the torque through an overall factor $g_0$.

\renewcommand{\theequation}{B\arabic{equation}}
\renewcommand{\thesection}{B}
\section{Asymptotic analysis in the high- and low-temperature regimes}
\label{appendixB}

In the high-temperature regime ($\xi\equiv\Om/\theta_0\ll1$, with $\theta_0=2\pi\kB T_0/\hbar$), the torque in Eq.~(\ref{M2bis}) admits the expansion
\begin{subequations}\label{asymptotics}
\begin{align}\label{largeOm}
M=-\frac{2}{189\pi}\,\frac{\hbar\,\theta_0^6\,\Om\,g_0}{c^6}\;
\Big[1 +\frac{14}{5}\,\xi^2 +\frac{112}{5\pi}\,\xi^5 +\frac{96}{5}\,\xi^6 +\mathcal{O}(\xi^7)\Big],\quad\quad\quad\quad\quad\quad\quad\;\,\xi\ll1,
\end{align}
so that the leading term is linear in $\Om$. In the opposite low-temperature regime ($\xi\gg1$), we find
\begin{align}\label{smallOm}
M=-\frac{128}{315\pi}\,\frac{\hbar\,\Om^7g_0}{c^6}\;
\Big[1+\frac{7}{64}\,\xi^{-4} -\frac{315\,\zeta(5)}{32\pi^5}\,\xi^{-5} +\frac{5}{128}\,\xi^{-6} +\mathcal{O}(\xi^{-7})\Big],\quad\quad\quad\quad\xi\gg1,
\end{align}
\end{subequations}
where $\zeta(5)\approx1.03693$. This expression shows a zero-temperature contribution scaling as $\propto\Om^7$. A self-contained derivation of Eqs.~(\ref{asymptotics}) is presented next.

We start from Eq.~(\ref{M2bis}), where we take $\Om>0$, approximate $g(\omega\pm\Om)\approx g_0$ (static limit), and use $n_{T_0}(\omega)=1/(\ee^{2\pi\omega/\theta_0}-1)$ with $\theta_0=2\pi\kB T_0/\hbar$. In what follows, we define $\xi=\Om/\theta_0$, $y=4\pi\xi$, and $x=2\pi\omega/\theta_0$. The $\omega$ integrals have dimensions of frequency to the seventh power, so we can recast the torque into
\begin{align}\label{MFy}
M=-\frac{\hbar\theta_0^7g_0}{288\pi^8c^6}\; F(y),
\end{align}
where
\begin{align}\label{Fy}
F(y)=&\int_0^y dx\,x^3(y-x)^3\big[1+n(x)+n(y-x)\big]
\\\nonumber
+ &\int_0^\infty dx\,x^3(y+x)^3\big[n(x)-n(y+x)\big]
\end{align}
is a dimensionless function written in terms of the Bose--Einstein function $n(x)=1/(\ee^x-1)$.

\subsection{Closed-form expression for the two-photon torque}

The zero-temperature part of the torque trivially contributes $\int_0^y dx\,x^3(y-x)^3=y^7/140$ to $F(y)$. In addition, the first integral of Eq.~(\ref{Fy}) can be simplified by noticing the result $\int_0^y dx\,x^3(y-x)^3 n(y-x)=\int_0^y dx\,x^3(y-x)^3 n(x)$, so we can write
\begin{align}\label{Fybis}
F(y)=\frac{y^7}{140}+2J(y)+K(y)
\end{align}
with
\begin{align}\label{Jy}
J(y)=\int_0^y dx\,\frac{x^3(y-x)^3}{\ee^x-1}
\end{align}
and
\begin{align}\label{Ky}
K(y)=\int_0^\infty dx\,x^3(y+x)^3\Big[\frac{1}{\ee^x-1}-\frac{1}{\ee^{x+y}-1}\Big].
\end{align}
Now, we apply the identities
\begin{subequations}\label{identities}
\begin{align}\label{ex1exp}
\frac{1}{\ee^x-1}=\sum_{\ell=1}^\infty \ee^{-\ell x}
\end{align}
and
\begin{align}\label{mlm1}
\int_0^\infty dx\,x^m \ee^{-\ell x}=\frac{m!}{\ell^{m+1}}.
\end{align}
\end{subequations}
In particular, Eq.~(\ref{ex1exp}) allows the expression in the squared brackets of Eq.~(\ref{Ky}) to be written as $\sum_{\ell=1}^\infty\left(1-\ee^{-\ell y}\right)\ee^{-\ell x}$. Together with the expansion $x^3(x+y)^3=x^6+3yx^5+3y^2x^4+y^3x^3$, this enables us to carry out the $x$ integral using Eq.~(\ref{mlm1}), yielding
\begin{align}\nonumber
&K(y)=
720\,\big[\zeta(7)-\textrm{Li}_7(\ee^{-y})\big]
+360\,y\big[\zeta(6)-\textrm{Li}_6(\ee^{-y})\big]
\\\label{Kyexact}
&+72\,y^2\big[\zeta(5)-\textrm{Li}_5(\ee^{-y})\big]
+6\,y^3\big[\zeta(4)-\textrm{Li}_4(\ee^{-y})\big].
\end{align}
Here, $\zeta(n)=\sum_{\ell=1}^\infty \ell^{-n}$ is the Riemann zeta function and $\textrm{Li}_n(x)=\sum_{\ell=1}^\infty \ell^{-n}\,x^\ell$ is a polylogarithm. For $J(y)$ [Eq.~(\ref{Jy})], we expand $x^3(y-x)^3=y^3x^3-3y^2x^4+3yx^5-x^6$ to obtain
\begin{subequations} \label{JyAm}
\begin{align} \label{Jyasym}
J(y)=y^3A_3(y)-3y^2A_4(y)+3yA_5(y)-A_6(y),
\end{align}
where
\begin{align}\label{Amy}
A_m(y)&=\int_0^y dx\,\frac{x^m}{\ee^x-1}
\\\nonumber
&=m!\Big[\zeta(m+1)-\sum_{j=0}^{m}\frac{y^j}{j!}\textrm{Li}_{m+1-j}(\ee^{-y})\Big].
\end{align}
\end{subequations}
The rightmost expression in Eq.~(\ref{Amy}) is obtained by applying Eq.~(\ref{ex1exp}) and integrating over $x$ by parts ($m$ times). Consequently, the full integrals in $F(y)$ [Eq.~(\ref{Fybis})] can be expressed in terms of polylogarithms and zeta functions.

\subsection{High-temperature limit: $\xi=\Omega/\theta_0\ll1$}

In $J(y)$ [Eq.~(\ref{Jy})], we have $x\le y\ll1$, so we can expand $1/(\ee^x-1)=1/x-1/2+x/12+\cdots$ and obtain
\begin{align}\nonumber
J(y)=\frac{y^6}{60}-\frac{y^7}{280}+\frac{y^8}{3360}+\cdots
\end{align}
by direct integration. For $K(y)$, expanding the exact polylogarithmic expression around $y=0$ gives
\begin{align}\nonumber
K(y)=\frac{16\pi^6}{21}y+\frac{2\pi^4}{15}y^3-\frac{y^6}{60}+\frac{y^7}{280}-\frac{y^8}{3360}+\cdots,
\end{align}
and therefore, Eq.~(\ref{Fybis}) leads to
\begin{align}\nonumber
F(y)=\frac{16\pi^6}{21}y+\frac{2\pi^4}{15}y^3+\frac{y^6}{60}+\frac{y^7}{280}+\frac{y^8}{3360}+\cdots,
\end{align}
which readily becomes Eq.~(\ref{largeOm}) by applying Eq.~(\ref{MFy}) and making the substitution $y=4\pi\xi$.

\subsection{Low-temperature limit: $\xi=\Omega/\theta_0\gg1$}

Now, we have $y\gg1$, so Eqs.~(\ref{JyAm}) become
\begin{align}\label{Jylow}
J(y)=&6\,\zeta(4)\,y^3-72\,\zeta(5)\,y^2
\\\nonumber
&+360\,\zeta(6)\,y-720\,\zeta(7)+\mathcal O(\ee^{-y}),
\end{align}
where we have used the scaling $\textrm{Li}_m(\ee^{-y})=\mathcal O(\ee^{-y})$. Likewise, Eq.~(\ref{Kyexact}) readily leads to
\begin{align}\label{Kylow}
K(y)=&6\,\zeta(4)\,y^3+72\,\zeta(5)\,y^2
\\\nonumber
&+360\,\zeta(6)\,y+720\,\zeta(7)+\mathcal O(\ee^{-y}).
\end{align}
Inserting Eqs.~(\ref{Jylow}) and (\ref{Kylow}) into Eq.~(\ref{Fybis}) and using the explicit values $\zeta(4)=\pi^4/90$ and $\zeta(6)=\pi^6/945$ for the zeta function, we find
\begin{align}\nonumber
F(y)=&\frac{y^7}{140}+\frac{\pi^4}{5}y^3-72\,\zeta(5)\,y^2
\\\nonumber
&+\frac{8\pi^6}{7}y-720\,\zeta(7)+\mathcal O(\ee^{-y}).
\end{align}
Finally, using Eq.~(\ref{MFy}) and setting $y=4\pi\xi$, we obtain Eq.~(\ref{smallOm}).

\renewcommand{\theequation}{C\arabic{equation}}
\renewcommand{\thesection}{C}
\section{Friction prefactor and spin-down time for a prolate diamond ellipsoid}
\label{appendixC}

We consider a prolate diamond ellipsoid with semi-major axis $R_x$ along $x$ and equal semi-minor axes $R_y=R_z<R_x$ along $y$ and $z$. The particle rotates about the $z$ axis. The polarizabilities entering Eq.~(\ref{gw}) have the analytical form \cite{O1945}
\begin{align}\nonumber
\alpha_{aa}=\frac{R_xR_y^2}{3}\frac{\epsilon-1}{1+(\epsilon-1)L_a},
\end{align}
along each Cartesian direction $a$. Here,
\begin{align}\nonumber
&L_x=\Big(\frac{1}{e^2}-1\Big) \bigg[\frac{1}{2e}\ln\Big(\frac{1+e}{1-e}\Big)-1\bigg],
\\\nonumber
&L_y=L_z=(1-L_x)/2
\end{align}
are the depolarization factors, expressed in terms of the eccentricity $e=\sqrt{1-R_y^2/R_x^2}$. We take $\epsilon=5.7$ as the static permittivity of diamond \cite{DNC98}. The off-diagonal components of the polarizability tensor vanish by symmetry. Substituting these expressions into Eq.~(\ref{gw}), we obtain $g_0=\eta R_x^6$, where $\eta$ depends only on the aspect ratio $R_y/R_x$. To maximize the frictional torque, we choose the value $\eta\approx0.48$, which occurs for $R_y/R_x\approx0.76$. For $R_x=100$~nm and $R_y=76$~nm, this gives $g_0\approx4.8\times10^{-31}$\,cm$^6$, which is the value used in Fig.~3 of the main text. For a rotation frequency $\Om/2\pi=6\,$GHz, the torque obtained from Eq.~(\ref{M2bis}) is $|M|\approx2.0\times10^{-33}$~N m. The moment of inertia of the particle is $I=(4\pi/15)\rho R_xR_y^2(R_x^2+R_y^2)\approx2.7\times10^{-25}$~g cm$^2$, where $\rho=3.51$~g/cm$^3$ is the density of diamond. The corresponding spin-down time is therefore $I\Om/|M|\approx5\times10^{11}$~s.

\renewcommand{\theequation}{D\arabic{equation}}
\renewcommand{\thesection}{D}
\section{Theory of rotational friction for a lossy particle}
\label{appendixD}

The theory developed in Sec.~\ref{appendixA} assumes a rotation frequency smaller than the internal excitation gap ($\Om<\eps_g$), and neglects direct optical absorption from the thermal vacuum bath by requiring $\eps_g-\Om\gg\theta_0$. When either condition is relaxed, friction can already arise from first-order transitions. This regime has been studied previously \cite{paper157,paper166}. Here, we revisit it with emphasis on its application to gapped particles.

From the theoretical framework formulated in Sec.~\ref{appendixA}, taking $\Om>0$, we obtain the first-order torque
\begin{align}\label{M11}
M=\frac{2\pi}{\hbar}\sum_{ij}\sum_\pm(\pm1)\big|\Delta^\pm_{ij}\big|^2\,\Big\{
&\big[n_{T_0}(\omega_i)+1\big]\, \big[n_{T_1}(\eps_j)+1\big]\, \delta(\eps_j+\omega_i\pm\Om)
\\\nonumber
+&n_{T_0}(\omega_i)\big[n_{T_1}\, (\eps_j)+1\big]\, \delta(\eps_j-\omega_i\pm\Om)
\\\nonumber
+&\big[n_{T_0}(\omega_i)+1\big]\, n_{T_1}(\eps_j)\, \delta(-\eps_j+\omega_i\pm\Om)
\\\nonumber
+&n_{T_0}(\omega_i)\, n_{T_1}(\eps_j)\, \delta(-\eps_j-\omega_i\pm\Om)
\Big\},
\end{align}
which consists of four terms corresponding to the possible combinations of photon absorption or emission (sum over $i$) and excitation or de-excitation of an internal particle mode (sum over $j$). We also introduce a particle temperature $T_1$, which determines the occupation of the internal modes through the Bose--Einstein distribution
\begin{align}\nonumber
n_{T_1}(\eps_j)=\frac{1}{\ee^{\hbar\eps_j/\kB T_1}-1}.
\end{align}
Using the prescription in Eq.~(\ref{sumi}) and carrying out the sums over photon polarizations and propagation directions according to Eq.~(\ref{sumsigma}), Eq.~(\ref{M11}) becomes
\begin{align}\nonumber
M=-\frac{2}{3c^3}\int_0^\infty \omega^3d\omega\, \sum_j&\big(p'^2_{j,x}+p'^2_{j,y}\big)\,\Big\{
\big[n_{T_1}(\omega+\Om)-n_{T_0}(\omega)\big]\, \delta(\eps_j-\omega-\Om)
\\\nonumber
&-\big[n_{T_1}(\omega-\Om)-n_{T_0}(\omega)\big]\, 
\big[\delta(\eps_j-\omega+\Om)-\delta(\eps_j+\omega-\Om)\big]
\Big\}.
\end{align}
The sum over internal modes can then be expressed in terms of the polarizability [Eq.~(\ref{alpha})], yielding
\begin{subequations}
\begin{align}\label{M1last}
M\!=\!-\frac{4\hbar}{3\pi c^3}\int_{-\infty}^\infty \!\!(\omega+\Om)^3d\omega\,
&\big[n_{T_1}(\omega)-n_{T_0}(\omega+\Om)\big]\, h(\omega),
\end{align}
where
\begin{align}\nonumber
h(\omega)=\frac{1}{2}\,{\rm Im}\{\alpha_{xx}(\omega)+\alpha_{yy}(\omega)\}
\end{align}
is an antisymmetric function [$h(\omega)=-h(-\omega)$]. Repeating this analysis to calculate the radiated power $P^{\rm rad}$, we obtain an expression analogous to Eq.~(\ref{M1last}), but with the leading factor $(\omega+\Om)^3$ in the integrand replaced by $-(\omega+\Om)^4$ \cite{paper157}. Then, energy conservation requires $-M\Omega=P^{\rm rad}+P^{\rm abs}$, where the left-hand side is the mechanical power lost by the rotating particle and the right-hand side is the sum of radiated and particle-absorbed powers. Combining these expressions, we find
\begin{align}\label{Pabslast}
P^{\rm abs}=&-\frac{4\hbar}{3\pi c^3}\int_{-\infty}^\infty (\omega+\Om)^3\omega\, d\omega\,
\big[n_{T_1}(\omega)-n_{T_0}(\omega+\Om)\big]\, h(\omega)
\\\nonumber
&-\frac{2\hbar}{3\pi c^3}\int_{-\infty}^\infty \omega^4\, d\omega\,
\big[n_{T_1}(\omega)-n_{T_0}(\omega)\big]\, {\rm Im}\{\alpha_{zz}(\omega)\}.
\end{align}
\end{subequations}
In the main text, we use these expressions to evaluate the particle--vacuum temperature difference and its contribution to rotational friction in gapped particles.

\renewcommand{\theequation}{E\arabic{equation}}
\renewcommand{\thesection}{E}
\section{Thermal friction of an axisymmetric particle with an internal energy gap}
\label{appendixE}

As shown in the main text, an axisymmetric particle with an internal gap frequency exceeding the rotation frequency ($\eps_g>\Om$) cannot experience frictional torque through processes that leave it in the internal ground state. Two-photon emission and higher-order photon processes are therefore forbidden. However, at finite temperature, the thermal population of photons with frequencies above the gap is nonzero, enabling friction through inelastic single-photon absorption followed by inelastic emission. Here, we estimate the resulting torque by approximating Eq.~(\ref{M1last}) under the assumptions $\Om\ll\eps_g$, $\theta_0\ll\eps_g$, and $h(\omega)=0$ for $|\omega|<\eps_g$.

We define $\alpha=T_1/T_0$, so that $n_{T_1}(\omega)=n_{T_0}(\omega/\alpha)$.

\subsection{Balance equation and $T_1/T_0$}

For simplicity, we neglect the second line of Eq.~(\ref{Pabslast}), corresponding to absorption by polarization along the rotation axis, and thus focus on disk-like particles. In any case, the temperature ratio scales as $\Omega^2$ (see below), so it does not affect the low-velocity torque of gapped particles with large $\epsilon_g$.

We first determine the ratio $T_1/T_0$ for which $P^{\rm abs}=0$ [see Eq.~(\ref{Pabslast})], corresponding to dynamical thermal equilibrium. For simplicity, we neglect the second line of Eq.~(\ref{Pabslast}), corresponding to absorption by polarization along the rotation axis, and thus focus on disk-like particles. In any case, the temperature ratio scales as $\Omega^2$ (see below), so it does not affect the low-velocity torque of gapped particles with large $\eps_g$. Separating the remaining integral into positive and negative $\omega$ regions with $|\omega|>\eps_g$, and using the asymptotic approximation $\int_{\eps_g}^\infty d\omega\,\ee^{-x\omega}f(\omega)\approx\ee^{-x\eps_g}f(\eps_g)/x$ for $x\gg1$ and smooth $f(\omega)$, we find
\begin{align}\nonumber
(\eps_g+\Om)^3 \Big[\alpha\ee^{-\hbar\eps_g/\alpha\kB T_0}-\ee^{-\hbar(\eps_g+\Om)/\kB T_0}\Big]
+(\eps_g-\Om)^3 \Big[\alpha\ee^{-\hbar\eps_g/\alpha\kB T_0}-\ee^{-\hbar(\eps_g-\Om)/\kB T_0}\Big]\approx0,
\end{align}
where we have used $h(-\omega)=-h(\omega)$ and $n_{T_0}(-\omega)=-1-n_{T_0}(\omega)$. This condition can be readily recast as
\begin{align}
\alpha\ee^{-\hbar\eps_g/\alpha\kB T_0}=\ee^{-\hbar\eps_g/\kB T_0}\frac{(1+\Om/\eps_g)^3\ee^{-\hbar\Om/\kB T_0}+(1-\Om/\eps_g)^3\ee^{\hbar\Om/\kB T_0}}
{(1+\Om/\eps_g)^3+(1-\Om/\eps_g)^3},
\label{alphacondition}
\end{align}
or equivalently,
\begin{align}\nonumber
\frac{T_1}{T_0}\approx
\frac{\hbar\eps_g/\kB T_0}
{W\bigg[
(\hbar\eps_g/\kB T_0)\ee^{\hbar\eps_g/\kB T_0}
\frac{(1+\Om/\eps_g)^3+(1-\Om/\eps_g)^3}
{(1+\Om/\eps_g)^3\ee^{-\hbar\Om/\kB T_0}+(1-\Om/\eps_g)^3\ee^{\hbar\Om/\kB T_0}}
\bigg]},
\end{align}
where $W$ is the Lambert function, defined implicitly by $W(x)\exp\{W(x)\}=x$. Retaining only the leading terms in $\Om/\eps_g$ and $\theta_0/\eps_g$, we find
\begin{align}\nonumber
\frac{T_1}{T_0}\approx
1+\frac{\kB T_0}{\hbar\eps_g}\ln\big[\cosh(\hbar\Om/\kB T_0)\big]
-\frac{3\,\Om}{\eps_g}\tanh(\hbar\Om/\kB T_0),
\end{align}
which is valid for arbitrary values of $\Om/\theta_0$.

\subsection{Threshold approximation for $M$}

Applying the same approximations as above, Eq.~(\ref{M1last}) becomes
\begin{align}\nonumber
M\approx-\frac{4\kB T_0}{3\pi c^3}h(\eps_g^+) \Big\{
&(\eps_g+\Om)^3 \big[\alpha\ee^{-\hbar\eps_g/\alpha\kB T_0}-\ee^{-\hbar(\eps_g+\Om)/\kB T_0}\big]
\\\nonumber
+&(\eps_g-\Om)^3 \big[\ee^{-\hbar(\eps_g-\Om)/\kB T_0}-\alpha\ee^{-\hbar\eps_g/\alpha\kB T_0}\big]
\Big\},
\end{align}
where $h(\eps_g^+)$ denotes the value of $h(\omega)$ immediately above the threshold $\omega=\eps_g$. Using Eq.~\eqref{alphacondition}, this expression reduces to
\begin{align}\nonumber
M\approx-\frac{8\kB T_0}{3\pi c^3}
\eps_g^3 h(\eps_g^+)\ee^{-\hbar\eps_g/\kB T_0}
\frac{\big[1-(\Om/\eps_g)^2\big]^3}{1+3(\Om/\eps_g)^2}
\sinh\Big(\frac{\hbar\Om}{\kB T_0}\Big).
\end{align}
To leading order in $\Om/\eps_g$, this becomes
\begin{align}\nonumber
M\approx-\frac{8\kB T_0\,\eps_g^3}{3\pi c^3} h(\eps_g^+)\,\ee^{-\hbar\eps_g/\kB T_0}
\sinh\Big(\frac{\hbar\Om}{\kB T_0}\Big)
\end{align}
and, if $\Om\ll\theta_0$, further reduces to
\begin{align}\label{Mapproxlast}
M\approx-\frac{8\hbar\eps_g^3}{3\pi c^3} h(\eps_g^+)\,\Om\,\ee^{-\hbar\eps_g/\kB T_0}.
\end{align}
We discuss Eq.~(\ref{Mapproxlast}) in the main text.

\end{widetext}


%

\end{document}